# Feature-Based Adaptive Tolerance Tree (FATT): An Efficient Indexing Technique for Content-Based Image Retrieval Using Wavelet Transform

Dr.P.AnandhaKumar
Department of Information Technology
Madras Institute of Technology Anna University Chennai,
Chennai,India
Email ID : annauniv@edu

V.Balamurugan
Research Scholar, Department of Information Technology
Madras Institute of Technology Anna University, Chennai
Chennai India
Email ID: vbalamurugan19@gmail.com

*Abstract*—**This paper introduces a novel indexing and access method, called Feature- Based Adaptive Tolerance Tree (FATT), using wavelet transform is proposed to organize large image data sets efficiently and to support popular image access mechanisms like Content Based Image Retrieval (CBIR).Conventional database systems are designed for managing textual and numerical data and retrieving such data is often based on simple comparisons of text or numerical values. However, this method is no longer adequate for images, since the digital presentation of images does not convey the reality of images. Retrieval of images become difficult when the database is very large. This paper addresses such problems and presents a novel indexing technique, Feature Based Adaptive Tolerance Tree (FATT), which is designed to bring an effective solution especially for indexing large databases. The proposed indexing scheme is then used along with a query by image content, in order to achieve the ultimate goal from the user point of view that is retrieval of all relevant images. Experimental results show that the FATT outperforms M-tree upto 200%, Slim-tree up to 120% and HCT upto 89%. FATT indexing technique is adopted to increase the efficiently of data storage and retrieval.**

## INTRODUCTION

One of the challenges in the development of a content based indexing and retrieval application is to achieve an efficient indexing scheme. Efficient image indexing and accessing tools are of paramount importance in order to fully utilize the increasing amount of digital data available on the internet and digital libraries. Recent years have been a growing interest in developing effective methods for indexing large image databases. Image databases are utilized in diverse areas such as advertising, medicine, entertainment, crime detection, education, military and in the domain of museum and gallery image collections etc. This motivates the development of efficient image indexing and retrieval systems and algorithms. However as the database grow larger, the traditional keywords based method to retrieve a particular image becomes inefficient[21],[26],[33],[35].

Content- based image retrieval (CBIR) approach has emerged as a promising alternative. CBIR has been a particular challenge as the image content covers a vast range of subjects and requirements from end users are often very loosely defined. In CBIR, images are indexed by its own visual contents, such as color, texture and shape. The challenge in CBIR is to develop the methods that will increase the retrieval accuracy and reduce the retrieval time. Hence, the need of an efficient indexing structure for images supporting popular access mechanisms like CBIR arises.[26], [33], [35].

In order to overcome such problems several content based indexing and retrieval applications have been developed such as the MUVIS system [22],[23],[29], photobook [31],VisualSEEK[37], Virage[40], and VideoQ[9] all of which are designed to bring a framework structure for handling and especially retrieval of the digital multimedia items, such as images, audio, and/or video clips. In this way similarity between two database images can be retrieved by calculating the similarity distance between their feature vectors. This is the general query by example (QBE) scenario, which on the other hand is expensive and CPU intensive especially for large databases. The main challenge of implementing such an index structure is to make it capable of handling high level image relationships easily and efficiently during access[7], [14],[21]

The indexing techniques can be mainly grouped in two types (1) Spatial Access Methods (SAMs) and (2) Metric Access Methods (MAMs) are proposed[21] .Initial attempts of SAMs are such as KD-trees[1],R-trees[16],SS-tree[41],Hybrid tree[8]etc. Especially for content-based indexing and retrieval in large scale databases. SAMs have several drawbacks and limitations. SAM-based indexing technique partitions and works over a single feature space.SAMs while providing good results on low dimensional feature space and do not scale up well to high dimensional spaces. SAM-based indexing schemes even becomes less efficient than sequential indexing for high dimensions.(2)Static and Dynamic Access Methods(MAMs)[17] are proposed such as VP-tree[43],MVP-tree[46],GNAT[6] whereas the dynamic ones, M-tree[7] and other M-tree variants such as M+-tree. The existing multidimensional index structures support CBIR by translating the content-similarity measurement into features level equivalence, which is a very difficult job and can result in erroneous interpretation of user's perception of similarity.

The indexing structures so far addressed are all designed to speed up any Query By Example(QBE)process by some multidimensional index structure.However all of them have significant drawbacks and shortcomings for indexing of large scale databases[1],[6],[7] ,[21], [41].In order to overcome such problems and provide efficient solutions to the aforementioned shortcomings of the indexing algorithm especially for image databases,we develop a MAM-based balanced and dynamic indexing technique called Feature-Based Adaptive tolerance Tree(FATT).

In this paper, we present FATT consists of root node(grandparent) is having the maximum equal parent nodes of 256 and depth of the tree depends upon the number of features considered and width of the tree depends upon the individual index value.

The rest of the paper is organized as follows: Section II presents the related work in the area of the indexing and retrieval. In Section III we introduce the general structure of FATT and algorithms. Section IV for wavelet transform and image coding. Section V presents the experimental results. Finally, Section VI concludes this paper.

### RELATED WORK

There are several multidimensional indexing techniques for capturing the low-level features like feature based or distance based techniques, each of which can be further classified as data-partitioned[3],[7],[16],[44]or space partitioned based algorithm [27],[34].Feature based indexing techniques project an image as a feature vector in a feature space and index the space. The basic feature based index structures are KD-tree [1], R-tree [16] etc.Researchers proposed several indexing techniques that are formed mostly in a hierarchical tree structure that is used to cluster the feature space. Initial attempts of such as KD-trees [1] used space –partitioning methods that divide the feature space into predefined hyper-planes regardless of the distribution of the feature vectors. Such regions are mutually disjoint and their union covers entire space. In R-tree [16] the feature space is divided according to the distribution of the database items and region overlapping may occur as a result. Both KD-tree and R-tree are the first examples of SAMs. Afterwards several enhanced SAMs have been proposed .R*-tree [2] provides consistently better performance by introducing a policy called "forced reinsert" than the R-tree and R+-tree[36].R*-tree also improves the node splitting policy of the R-tree by taking overlapping area and region parameters into consideration.Lin et al proposed TV-tree[19],which called as telescope vectors. These vectors can be dynamically shortened assuming that only dimensions with high variance dimensions can be neglected. Berchtold et al [3] introduced X-tree, which is particularly designed for indexing high dimensional data. X-tree avoids overlapping of  region bounding boxes in the directory structure by using a new organization of the directory as a result, X-tree outperforms both TV-tree and R*-tree significantly. It is 450 times faster than R-tree and between 12 times faster than the TV-tree when the dimension is higher than two and it also provides faster insertion times.Still, bounding rectangles can overlap in higher dimensions. In order to prevent this, White and Jain proposed the SS-tree [41], an alternative to the R-tree structure, which uses minimum bounding spheres instead of rectangles. Even though SS-tree outperforms R*-tree, the overlapping in the high dimensions still occur. Thereafter, several other SAM variants are proposed such as SR-tree[20],$S^2$-tree[42],Hybrid-tree[8],A-tree[38],IQ-tree[5],Pyramid   –tree[4],NB-tree[13] etc.The aforementioned degradations and shortcomings prevent a wide usage of SAM based indexing structures especially for large scale image collections. In order to provide a more general approach to similarity indexing for several MAM- based indexing techniques have been proposed.Yianilos[43] presented VP-tree that is based on partitioning the feature vectors into two groups according to their similarity distances with respect to a reference point, a so-called vantage point.Bozkaya and Ozsoyoglu[47] proposed extension of the VP-tree, a so-called MVP-tree(multiple vantage point),which basically assigns m vantage points to a node with a fan out of $m^2$ .They reported 20% -80% reduction of similarity distance computation compared to VP-trees. Brin [6]introduced the geometric near-neighbor access tree(GNAT) indexing structure,which chooses k split points at the top level and each of the remaining feature vectors are associated with the closest split points. GNAT is then built recursively and the parameter k is chosen to be a different value for each feature set depending on its cardinality.Koikkalainen and Oja introduced TS-SOM [25] that is used in PicSOM [28] as a CBIR indexing structure.TS-SOM provides a tree –structure vector quantatization algorithm. Other SOM-based approaches are introduced by Zhang and Zhong [45],and Sethi and Coman[39].All SOM-based indexing method rely on training of the levels has a pre-fixed node size that has to arranged according to the size of the database. This brings a significant limitation ,that is they are all static indexing structures,which do not allow dynamic construction or updates for particular database. Retraining and costly reorganizations are required each time the content of the database changes (i.e., new insertions and deletions),that is indeed nothing but rebuilding the whole indexing structure from the scratch. Similarly, the rest of the MAMs, so far addressed present several shortcomings. Contrary to SAMs, these metric trees are designed only to reduce the number of similarity distance computations, paying no attention to I/O costs (disk page accesses).They are also intrinsically static methods in the sense that the tree structure is built once and new insertions are not supported.Furthermore, all of them build the indexing structure from top to bottom and hence the resulting tree is not guaranteed to be balanced. Ciaccia et al [7] proposed the M-tree to overcome such problems. The M-tree is a balanced and dynamic tree, which is built from bottom to top, creating a new level only when necessary. The node size is a fixed number M, and therefore the tree height depends on M and the database size. Its performance optimization concerns both CPU computational time for similarity distances and I/O costs for disk accesses for feature vectors of the database items. Traina et al [15] proposed Slim- tree, an enhanced variant of M-trees, which is designed for improving the performance by minimizing the overlaps between nodes. They introduced two parameters, "fat-factor" and "blot factor", to measure the degree of overlap and proposed the usage of minimum spanning tree (MST)[24],[32],for splitting the node. Another slightly enhanced M-tree structure, a so-called M+-tree, can be found.[46].Serkan kiranyaz and Moncef Gabbouj [21] proposed Hierarchical cellular tree(HCT) has no limit for cell size as long as cell keeps a definite compactness.Any action (i.e,an item insertion)it consequent reactions cannot last indefinitely due to the fact that each of them can occur only in a higher level and any HCT body has naturally limited number of levels. Ming Zhang and John Fulcher[48] proposed GAT-trees which is neural network based adaptive tree but it takes longer time to train the network and it gives accuracy 5% higher than general tree random noise is added, 7% accuracy when gamma value of Gaussian noise exceeds 0.3.

1. FATT indexing is MAM-based and have a hierarchical structure, i.e., levels *(m)*.

2. FATT designed, dynamically and balanced in a top-to-bottom fashion to overcome the problem of overlap between nodes in metric spaces.

3. Root node (grant parent(*A*)) grows from the root node to leaves, each parent node is represented with several left child*(lchild)* and rightchild*(rchild)* nodes and maximum of 256(*N*) nodes.

4. Each child node immediately proceeding with equal left child*(lchild)* and rightchild*(rchild)*.For both *lchild* and *rchild* of *A+1* is at *NA+N=N(N+A)*.Unless *N(A+1)>m* in which case (*A+1*) has no *lchild*.Hence this representation can clearly be used for all image indexing and fast searching process.

FATT STRUCTURE AND ALGORITHMS

*FATT Structure*

FATT is balanced, adaptive and an indexing tree .It is designed in a complete *N*-ary, where *N* is the number of nodes. The FATT is constructed based on the index code developed. The height of the tree depends upon the number of features considered and width of the tree depends upon the number individual index codes that are used in code generation. The leaf node contains the index value. The retrieval performance depends upon the tree construction. The size of the tree can be changed depending upon the number of images that has to be stored. The structure of the tree can be changed depending upon images found. The general structure *(4 levels, m =4, N= 256)*of the FATT is shown in Fig.1.The maximum number unique index value for a *m*-levels, *N*-nodes tree is given as $=N^m$

No. of nodes at top level *(m)*=1
No. of nodes for each node at first level *(m₁)* =256
No. of nodes for each node at second level *(m₂)* =256
No. of nodes for each node at third level *(m₃)* =256
No. of nodes for each node at fourth level *(m₄)* =256
No. of individual index value =$(1\times256\times256\times256\times256)^4$=$4^{10}$

Where *N=256, m=4*

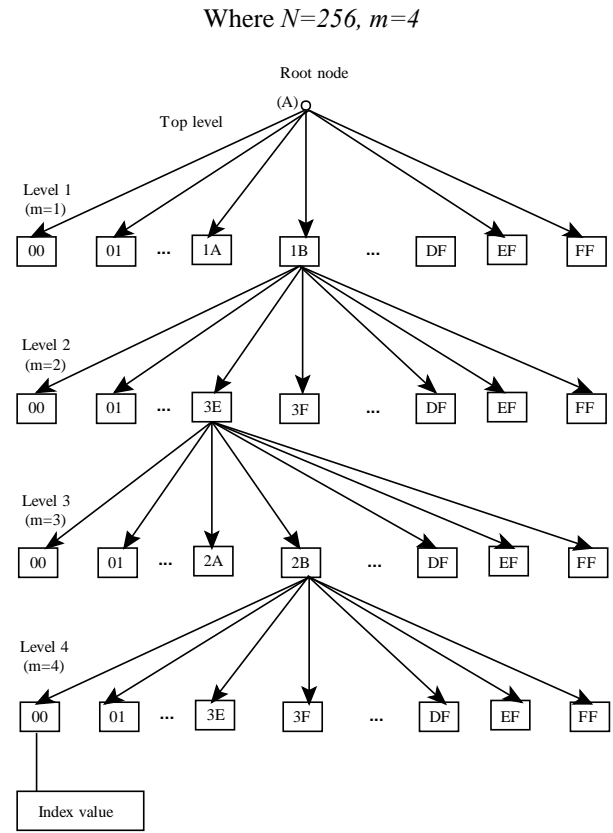

Fig.1.General structure of the FATT(*N=256, m=4*)

*FATT Algorithms and Operations*

There are 3 FATT algorithms and operations (i.e) image insertion, searching and retrieval (indexing).

The insertion algorithm Insert *(nextImage,levelNo)* first performs the searching algorithm which hierarchically FATT from root node(grant parent) to the least child node .In order to locate most suitable image for *nextImage*.Once the least node is located,the image is inserted into the immediate parent node. Let the *nextImage* be the image to be inserted into a target level indicated by *levelNo*. Accordingly the insertion algorithm as follows and the sample insertion of 2 nodes ,4 levels,( *N=2, m=4)* 3 nodes 4 levels(*N=3, m=4)* is shown in Fig.2. and Fig.3

*1.Insertion Algorithm*

Insert*(nextImage,levelNo)*
Let the root level number; *root LevelNo* and the single image in the top level:

➤ If*(levelNo>rootlevelNo)* then do;
Append *nextImage* into least *lchild* node
➤ If*(levelNo=rootlevelNo)* then do;
Assign *rootlevelNo=LevelNo*
➤ If*(levelNo<root levelNo)* then do;
Append *nextImage* into least *rchild* node.

Check least child node for post- processing
> If leastchild node is split then do;
Create the least *lchild* node equal to immediate parent node
> Insert(*nextImage*,immediate *root levelNo)*
Else if insertion of least child node then do;
Let least *lchild* and *nextImage* be parent and new child node
Else if unchange the immediate parent node
Return.
*A.Insertion of 4-levels and 2-nodes (m=4,N=2)*
Number of levels, *m* =4
Number of nodes, *N* =2
Then the total number of index values $(N^m)$ is16

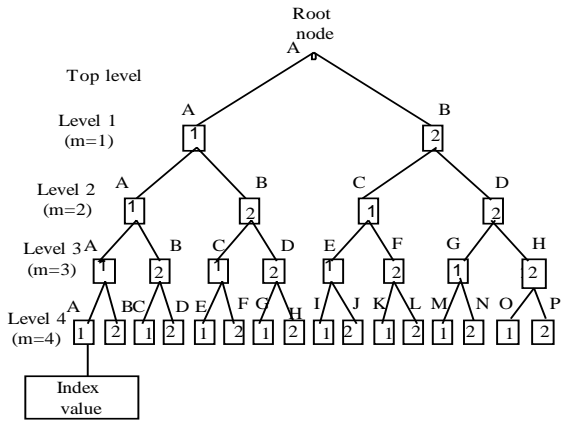

Fig.2.A sample FATT *(m=4,N=2)*

If a complete FATT with *N* =2, *m* =4 depth first search is represented sequentially as above then for then for grand parent node with index *A, 1≤A≤m* we have,

Case (1) grandparent *(A)* is *at (A/2)* if *A* =1 when *A=1, A* is the root and has no parents.

Case (2) leftchild *(A)* is at *2A if 2A≤m*.If *2A>m*, then grand parent *A* has no left child(*lchild*)

Case (3) rightchild *(A)* is at *2A+1 if 2A+1≤m*,If *2A+1>m*, then *A* has no rightchild(*rchild*)

Proof:

We prove case(2) and case(3) is immediate consequence of case(2) and the numbering nodes on the same level from left to right case (1) follows from case (2) and case (3).We prove case (2) by induction on *A*..For *A* =1,clearly the *lchild* is at *2* unless *2>m* in which case (1) has no *lchild*.Then two nodes immediately proceeding *lchild(A+1)*.In representations are the *rchild* of *A* and *lchild* of *A*.The *lchild* of *A* is at *2A*.Hence $2A+$ the *lchild* of *A+1* is at *2A+2=2(A+1)*.Unless *2A+1>m*, in which case *A+1,*has no *lchild*.

*B.Insertion of 4-levels and 3-nodes(m=4,N=3)*

Number of levels, *m* ==4
Number of node, *N* =3
Then the total number of index values $(N^m)$ is 81

If a complete FATT with *N=3,m=4* the depth first search (i.e.,*DFS*) is represented sequentially as above then for grand parent node with index *A , 1≤A≤m* we have,

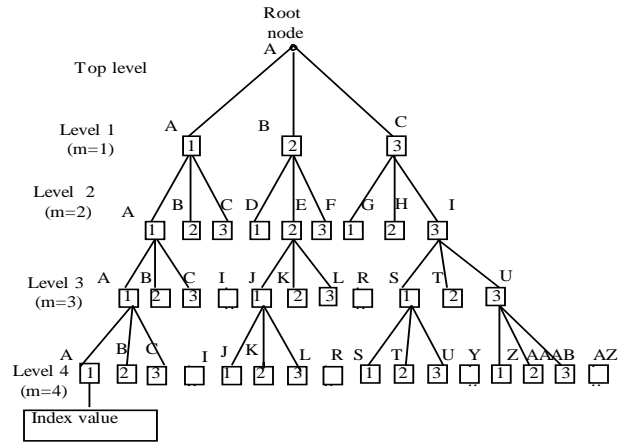

Fig.2

Fig.3..A sample FATT *(m=4,N=3)*

Case (1) grand parent *(A)* is at *(A/3)if A≠1*.When *A=1, A* is the root and has no parent

Case (2) leftchild *(A)* is at *3A if 3A≤m*.If *3A>m*,the *A* has no left child(*lchild*)

Case (3) middlechild*(A)* is at *3A+1 if 3A+1≤m*.If *3A+1>m*, then *A* has no middle child(*mchild*)

Case (4) rightchild *(A)* is at *3A+2 if 3A+2≤m*.If *3A+2>m*, then then *A* has no right child(*rchild*)

Proof:

We prove case(1),(3) and (4) is an immediate consequence of case (2) and numbering of nodes on the same level left to middle and right.Case(1)follows (2),(3) and (4).We prove case (2) by induction on *A*.For *A=1,*Clearly *lchild* is at *3* unless *3>m*,then nodes immediately proceeding *lchild(A+1),* in the representations are the middle child of *A* and the *lchild(A).* The *lchild(A)* is at *3(A),* hence the *lchild* of *3A+2* unless *3(A+1)>m..*

In general if a complete FATT with number of maximum nodes is 256(*N=256*) (ie,depth =$N^m$ is represented sequentially as above then for grandparent node with index *A,1≤A≤N*.we have,

Case (1) grandparent *(A)* is at *(A/N)* if *A≠1* when *A=1,A* is the root and no parent.

Case (2) leftchild *(A)* is at *2A* if *2A ≤256*.If *2A>256* then *A* has no *lchild. mchild(A)* at *2A+1* if *2A+1≤256.*If *2A+1≤256.*If *2A+1>256* then *A* has no *mchild.*

Case   Case (3) rightchild *(A* ) is at *2A+2* if *2A+2≤256.*
If *2A+2>256* then *A* has no   *rchild* .we prove that case(2),(3)and (4) is an immediate consequence of case (2) and the numbering of nodes on the same level from left to right case(1) follows from case (2),(3) and (4).we prove case (2) by induction on *A* at *2.* Unless *2>256* in which case (1) has no *lchild* .Then,the two nodes immediately proceeding *lchild (A+1)* in the representations are the *mchild A* and the *lchild* of *A*.The *lchild* of *A* is at *2A* .Hence the *lchild* of *A+1* is at *2A+2=2(A+1)* unless *2(A+1)>256* in which case *A+1* has no *lchild.*

## 2. Searching Algorithm

Depth first search algorithm is used for traversing the tree. In this algorithm search as deeply as possible by visiting a node, and then recursively performing depth –first search on each adjacent node.

Search*(nextImage,levelNo)*
➤ If the *parentnode >root node* then do;
Go to the *lchild* of the *root node*
Until no *lchild* is avilable
Check whether the indexing value is obtained
➤ Elseif *parentnode< root node*
Return the index value
➤ Else not found traverse step 1 and 2
Repeat the steps 1,2 and 3 until index value is found.

The searching algorithm search*(nextImage,levelNo)* first perform the searching with *root node.*
Case (1) If *parentnode >root node* searching beginning with *lchild* node until no leftmost child is available.
Case (2) If *parentnode< root node* then return the index value otherwise search for leftmost child node is available

## 3. Retrieval Algorithm

FATT can index large image-scale database using index code generated and euclidean similarity measure used as distance metric

Index*(nextImage,levelNo)*
➤ If the *parentnode >root node* then do;
Go to the *lchild* of the *root node*
Until no *lchild* is avilable
Check whether the indexing value is obtained
➤ Elseif *parentnode< root node*
Return the index value
➤ Else not found traverse step 1 and 2
➤ Repeat the steps 1,2 and 3 until index value is found.

The indexing algorithm index*(nextImage,levelNo)* first perform the indexing with *root node*

Case (1) If *parentnode >root node* indexing beginning with *lchild* node until no leftmost child is available.Check whether index value is available.
Case (2)If *parentnode< root node* then return the index value otherwise index for leftmost child node is available

WAVELET TRANSFORM AND IMAGE CODING

### Wavelet Transform

A 2 dimensional-discrete wavelet transform(2D-DWT) is a process which decomposes a signal, that is a series of digital samples, by passing it through two filters, a low pass filter L and high pass filter H. The low pass sub band represents a down sampled version of the original signal. The high pass sub band represents residual information of the original signal, needed for the perfect reconstruction of the original set from the low resolution version.

Since image is typically a two dimensional signal, a 2D equivalent of the DWT is performed. This is achieved by first applying the L and H filters to the lines of samples, row by row, then refiltering the output to the columns by the same filters. As the result, the image shown in Fig.4 is divided into 4 subbands, LL, LH, HL and HH as depicted in Fig.5. The LL sub band contains the low pass information of horizontal, vertical and diagonal orientation. The LL sub band provides a half sized version of input image which can be transformed again to have more levels of resolution..Generally, an image is partitioned into L resolution levels by applying the 2D DWT (L-1) times.[11].

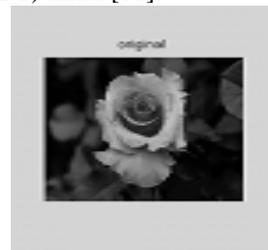 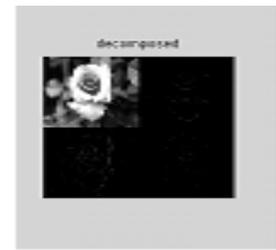

Fig.4.Orginal image      Fig.5.One level decomposition

By wavelet transform, we mean the decomposition of an image with family of real orthogonal bases $\psi_{m,n}($ obtained through translation and dilation of a kernel function $\psi($ known as mother wavelet.[11].

$$\psi_{m,n}(x) = 2^{\frac{-m}{2}}\psi(2^{-m}x -$$
(1)

Where and are integers. Due to the orthonormal property,the wavelet coefficients of a signal $f($ can be easily computed via

$$c_{m,n} = \int_{-\infty}^{+\infty} f(x)\,\psi_{m,n}(x),$$
(2)

and the synthesis formula

$$f(x) = \sum_{m,n} c_{m,n}\psi_{m,n}(.$$
(3)

can be used to recover $f($ from its wavelet coefficients

To construct the mother wavelet $\psi($ we may first determine a scaling function $\phi($which satisfies the two-scale difference equation

$$\phi(x) = \sqrt{2}\sum_{k} h(k)\phi(2x -$$
(4)

Then the wavelet kernel $\psi($ is related to the scaling function via,

$$\psi(x) = \sqrt{2} \ldots \tag{5}$$

where

$$g(k) = (-1)^k h(1 - \tag{6}$$

The coefficients      in (4) have to several conditions for the set of basis wavelet functions in (1) to be unique, orthonormal, and have a certain degree of regularity [11], [18].

The coefficients      and      play a very crucial role in a given discrete wavelet transform. To perform the wavelet transform does not require the explicit forms of      and

$\psi(x)$ but only depends on      and      .Consider a J-level

wavelet decomposition which can be written as

$$f_0(x)$$
$$= \sum_k (C_{j+1,}k\phi_{j+1,}\kappa(x) + \sum_{j=0}^j d_{j+1,}k\psi_{j+1,}k(x \tag{7}$$

Where coefficients      are given and coefficients      and      at scale      are related to the coefficients      at scale      via

$$c_{j+1,n} = \sum_k c_{j,k} h(k - 2 \tag{8}$$

$$d_{j+1,n} = \sum_k c_{j,k} g(k - 2n \tag{9}$$

Where  $0 \le j \le$ .Thus,(8) and (9) provides a recursive algorithm for wavelet decomposition through  $h($  and  $g($ ,and the coefficients  $c$  for a low resolution component  $\phi_{j,k}($ .By using a similar approach,we can derive a recursive algorithm for function synthesis based on its wavelet coefficients  $d$ ,  $0 \le j \le$  and  $c$ 

$$c_{j,k} = \sum_n c_{j+1,n} h(k - 2n) + \sum_n d_{j+1,n} g(k - 2 \tag{10}$$

It is convenient to view the decomposition (8) as passing a signal  $c$  through a pair of filters      and      with impulse responses  $\tilde{h}($  and  $\tilde{g}($  and down sampling the filtered signals by two (dropping every other sample), where  $\tilde{h}($  and  $\tilde{g}($  are defined as

$$\tilde{h}(n) = h(-n) \ldots$$

The pair of filters $H$ and $G$. correspond to the halfband lowpass and highpass filters,respectively,and are called the quadrature mirror filters in the signal processing literature[11] The reconstruction procedure is implemented by upsampling the subsignals      and      and filtering with      and      ,respectively, and adding these two filtered signals together.Usually the signal decomposition scheme is performed recursively to the output of lowpass filter      .

The wavelet packet basis functions      can be generated from a given function      as follows

$$W_{2n}(x) = \sqrt{2}\sum_k h(k_{\cdots} \tag{11}$$

$$W_{2n+1}(x) = \sqrt{2}\sum_k g(k) \ldots \tag{12}$$

Where the function      can be identified with the scaling function and      with the mother wavelet      .Then,the wavelet packet bases can be defined to be the collection of orthonormal bases composed of functions of the form      ,where *l*, *k*  ,*n*.  Each element is determined by a subset of the indices: a scaling parameter *l*, a localization parameter *k*, and an oscillation parameter *n*.

The 2D wavelet (or wavelet packet) basis functions can be expressed by tensor product of two 1-D wavelet (or wavelet packet) basis functions along the horizontal and vertical directions. The corresponding 2-D filter coefficients can be expressed as

$$\tag{13}$$

$$h_{HL}(k.l) = g(k)h(l), h_{HH}(k.l) = g(k)g(l) \tag{14}$$

Where the first and second subscripts in (13) and (14) denotes the low pass and highpass filtering characteristics in the  -and  -directions respectively.[11]We have applied orthogonal wavelet transformation with dyadic subsampling.Wavelet decomposition of images is performed using an db4 wavelet basis function. On application of the the above procedure,for an image of size 256 × 256, db4 wavelet yields suband matrices of 128×128 at the first level,64×64 at the second level, 32×32 at the third level, 16×16 at the fourth level of wavelet decomposition.

*Image Coding*

In image coding, a six digit code is generated for each image feature which is to be stored and retrieved .The code generated depends upon the value of the determinant generated. The number of different codes that can be generated depends upon the value of the determinant .The algorithm for finding the code is given below

Code=function (determinant value)

The code is generated depends upon the determinant value of the matrix. The function generates the code between 0 to F based upon the value of the determinant. It is a function in which the input to the function is the determinant value obtained. Based upon the range of the determinant obtained the function returns value which is the code for the image.The database to store the images is a single table with three fields.The three fields are serial number, image and index value.

*Image Coding Algorithm*

Step1:  Get the matrix of the individual feature ($m \times m$)
Step2:  Find the transpose of the matrix obtained
 ➤ Read the $m \times m$ matrix
 ➤ Give the input of the $m \times m$ matrix elements
 ➤ Transpose the $m \times m$ of the input matrix elements
 Step 3: Multiply the two matrices to get a square matrix.
 ➤ Read the order of $m \times m$ for the first matrix
 ➤ Read the order of $p \times q$ for the second matrix
 ➤ Give the input of the $m \times m$ of the first matrix elements
 ➤ Give the input of the $p \times q$ of the second matrix elements
 ➤ For i=0;i<m;i++,For j=0;j<q;j++,For k=0;k<p;k++
 ➤ $C[i,j]=C[i,j]+(A[i,k] \times B[k,j])$
Step4:  Get the determinant value of the matrix ($m \times m$)
 ➤ Read the $m \times m$ matrix
 ➤ Give the input $m \times m$ matrix elements
 ➤ The determinant value of the $m \times m$ matrix
 ➤ Determinant value given by $A[i,j] \times (A[i+1,j+1] \times A[i+2, j+2] - A[i+2, j+1] \times A[i+1,j+1]) A[i+1,j+1] \times (A[i+1,j] \times A[i+2, j+2] - A[i+2, j] \times A[i+1,j+2]) + A[i+1,j+2] \times (A[i+1,j] \times A[i+2, j+1] - A[i+1,j+1] \times A[i+2, j])$
Step5:  Based on the range of the determinant value the code generated.

*If x   0 and x≤10  then y=00*
*Else if x>10 and x≤20 then y=01*
*Else if x>20 and x≤30 then y=02*
*Else if x>30and x≤40 then y=03*
*Else if x>40 and x≤50 then y=04*
*Else if x>50 and x≤60 then y=05*
*Else if x>60and x≤70 then y=06*
*Else if x>70and x≤80 then y=07*
*Else if x>80 and x≤90 then y=08*
*Else if x>90 and x≤100 then y=09*
*Else if x>100 and x≤110 then y=10*
*Else if x>110 and x≤120 then y=11*
*Else if x>120 and x≤130 then y=12*
*Else if x>130 and x≤140 then y=13*
*Else if x>140 and x≤150 then y=14*
*Else if x>150 then y=15*
*Endif*
*.*
*.*
*Endif*

### D.  Error Handling Phase

Errors may occur during retrieval of images from the database. The error is caused by the code change during the generation purpose. The code generated is made up of six parts each obtained from each individual feature. The error occurs when determinant value obtained returns a code that differs from the code of the image stored in the database. This changes the code entirely. The severity of the error lies in the level it occurs. When the error occurs at the bottom most level the severity is minimum. Whenever such error occurs searching the neighboring nodes can rectify it. This only possible only if the error occurs at the last level. If the error occurs at higher level tracing the tree becomes difficult. Example
Let the image in the database have the node, 122300
Let the code generated for the image which is being given as the input be, 122301
Searching the adjacent nodes, which would return the image required, could rectify this error.

### EXPERIMENTAL RESULTS

We have tested the FATT to index and retrieve the images with three different databases through a query process such as query by example (QBE). Database 1-D1 consists of 1000 natural scene images of size 256×256 of 60 selected categories are from D1:various beach images, sunsets, waterfalls, clouds, mountains and glaciers etc from COREL. Database 2-D2consists of 10000 natural scene images of 100 categories similar contents with D1 from COREL. Database.Selected categories are rose,sunflower,dinosaur,,bus,lion,elephant etc.We implemented in MATLAB 7.4 under Windows vista. The experiments were performed on a Pentium Core 2 Duo,1.66GHz PC with 1GB RAM.Our implementations for node access ,distance computations and retrieval process.

One indexing and retrieval results from D1 shown in Fig.6 , D2 is shown in Fig.7 and for D3 shown in Fig.8.The effectiveness of the proposed tree have been validated by a large number of experiments.

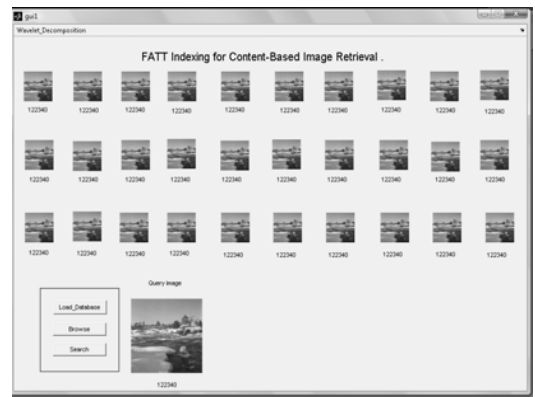

Fig.6.Retrieval of images from D1

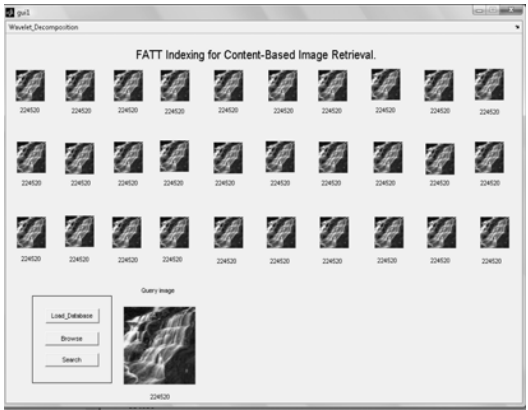

Fig.7.Retrieval of images from D2

Fig.8.Retrieval of images from D3

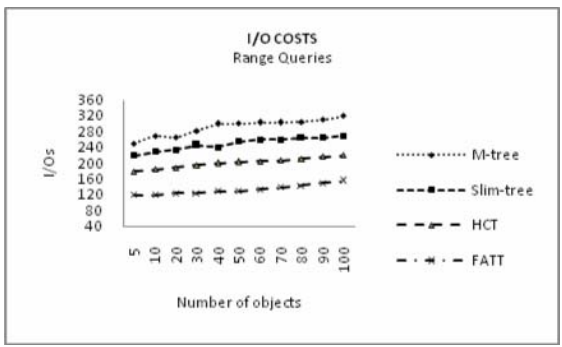

Fig.9.I/Os versus number of objects in D1

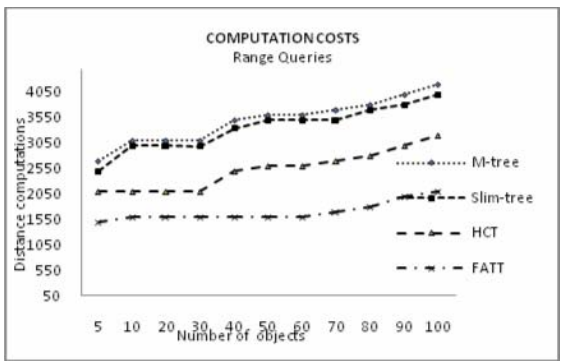

Fig.10.Computation costs versus no. of objects in D1

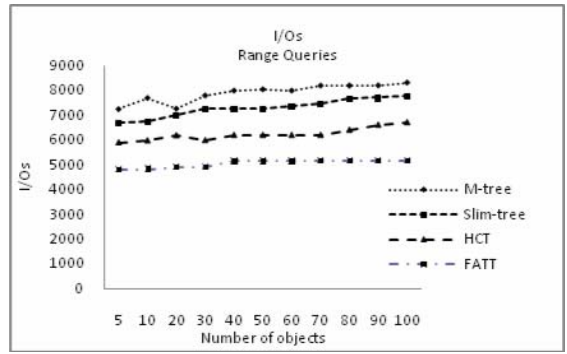

Fig.11.I/Os versus number of objects in D2

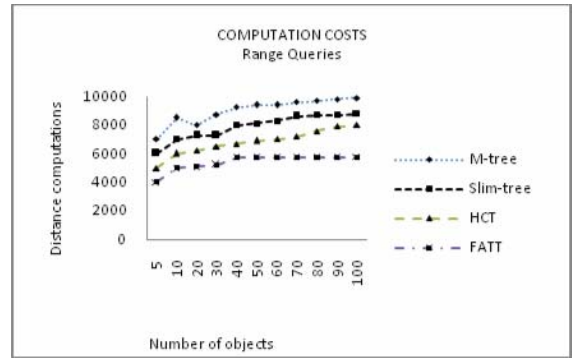

Fig.12.Computation costs versus no. of objects in D2

Since FATT is balanced and dynamic tree the performances like I/Os and computation costs is compared with existing balanced and dynamic tress such as M-tree, Slim-tree, HCT. The FATT consistently outperforms irrespective of the database size .Fig.9,Fig.11.,Fig.13 shows the I/Os versus number of objects in D1,D2 and D3.From the experimental though database size increased but the I/O cost maintains less and uniformly.upto 83% for the D1,upto81% for D2 and upto 82% for D3 the performance is achieved.

Fig.10. Fig.12, Fig.14.Shows the computation costs versus no.of objects for D1and D2.The computation cost the for D1decreased up to 82% up to 81% for D2 and upto83% for D3. Average distance computations performance also outperforms for D1 and D2.Experimental results show that the performance is no degradation irrespective of the database size is increased.FATT indexing technique can be conveniently used for retrieval and to achive the users goal to retrieve the most relevant images.

*Complexity of the algorithm*

Suppose we are given $n$ images A, B, C…Z and suppose the images are inserted in order into a FATT node. There are $n$ permutations are required for $n$ images. Each such permutations will give rise to a corresponding tree. It can be shown that the average depth of the $n$ tree is approximately $log^2 n + 1$,Accordingly ,the average running time $f(n)$ to search for image in FATT with n images is proportional to $\log_2 n$, (i.e,) $f(n) = O(\log^2 n + 1)$.

TABLE III
COMPLEXITY ANALYSIS

| $\log_2 n$ | $n$ | $n\log_2 n$ | $n^2$ | $n^3$ |
|---|---|---|---|---|
| 1 | 2 | 2 | 4 | 8 |
| 1.584 | 3 | 4.754 | 9 | 27 |
| 2 | 4 | 8 | 16 | 64 |
| 2.321 | 5 | 11.609 | 25 | 125 |
| 2.584 | 6 | 15.509 | 36 | 216 |
| 7.807 | 7 | 19.651 | 49 | 343 |
| 3 | 8 | 24 | 64 | 512 |
| 3.169 | 9 | 28.529 | 81 | 729 |
| 3.321 | 10 | 33.329 | 100 | 1000000 |
| 8 | 256 | 20.48 | 65536 | 16777216 |

FATT nodes represent the database images, this requires prior calculation of the relative similarity distances and hence yields a$O(n^2)$computational cost. The nodes are constructed dynamically since images can be any time inserted since such operation requires $O(n^3)$. Therefore, an incremental FATT algorithm is adaptive based on leaf nodes. This is a sequential algorithm and $O(n)$ computational complexity per image and hence $O(n^2)$ overall cost as desired.

CONCLUSION AND FUTURE WORK

In this paper, a novel indexing technique for content based image retrieval is designed particularly for fast insertion, searching and indexing, moreover to tackle the problem of overlap between the nodes. The M-tree has fixed cell policy due to its natural deficiency not suitable in dynamic environments and data are subject to permanent changes. Slim-tree is the extension of M-tree that speed up insertion using node splitting algorithm while also improving the storage utilization, unfortunately this algorithm does not guarantee the minimal occupation of each node. Whereas HCT has flexible cell policy but the levels are limited. Our proposed FATT greatly improves the performance comparing with M-tree, Slim-tree and HCT indexing schemes in I/Os costs, computation costs and average distance computations. The insertion speed of our FATT improves up to 84% compared with M-tree, up to 75% compared with Slim-tree, up to 70% compared with HCT. The searching speed of improves up to 146% compared with M-tree, up to 105% compared with Slim-tree, up to 89% compared with HCT. The possibility of using clustering, neural networks and other multimedia databases such as audio, video will also give attractive performances could be explored in future

P.AnandhaKumar received B.E degree  in Electronics and Communication from Government College of Engineering,Salem,Tamil nadu.  and M.E degree in Computer Science and Engineering from Government College of Technology, Coimbatore. He received PhD degree from College of Engineering Guindy in the Department of Computer Science and Engineering at Anna university,Chennai.Currently working as Assistant Professor in the Department of Information Technology at Madras Institute of Technology,Anna University- Chennai,India. He has published several papers in international, national journals and conferences. His areas of interests includes content-based image indexing techniques and frameworks, image processing and analysis,video analysis,fuzzy logic,pattern recognition,knowledge management and semantic analysis .

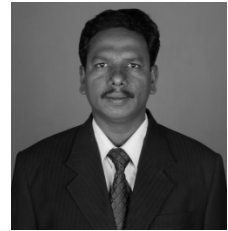

V.Balamurugan received B.E degree in Electrical and Electronics Engineering from Government College Technology, Coimbatore and M.E degree in Applied Electronics with distinction from Karunya Institute of Technology, Coimbatore. Currently pursuing PhD degree in the Department of Information Technology at Madras Institute ofTechnology, Anna University-Chennai. India. His current research interest includes in the area of content-based image retrieval, image processing and analysis, digital signal processing, wavelet theory and application.